\documentclass{optica-article}

\journal{opticajournal} 

\articletype{Research Article}

\usepackage{lineno}

\begin{document}

\title{Rydberg excitation through detuned microwave transition in rubidium}

\author{E. Brekke\authormark{1} and C. Umland}

\address{\authormark{1,*}St. Norbert College, Deparment of Physics, De Pere, WI 54115, USA\\}

\email{\authormark{*}erik.brekke@snc.edu} 


\begin{abstract*} 
We study the excitation of Rydberg states in warm rubidium vapor.  Using an inverted wavelength excitation scheme, we observe the effect of microwave coupling between Rydberg states through electromagnetically induced transparency.  We observe AC stark shifts of the Rydberg states from the microwave coupling, and demonstrate detuned excitation to a secondary Rydberg state.  These results show flexibility in excitation process and state selection necessary for a variety of wave-mixing processes using Rydberg states.


\end{abstract*}

\section{Introduction}
Rydberg atoms have continued to be pursued for their unique characteristics, including strong interactions, long lifetimes, and closely spaced levels with microwave transitions \cite{Gallagher_1994, Saffman:10}.  
Thermal vapor Rydberg atoms have the appeal of experimental simplicity, while still showing excellent characteristics for electromagnetically induced transparency (EIT) \cite {Mohapatra:07}, coherent excitation processes \cite {Kwak:16, Jin:22},  four-wave mixing \cite {Kolle:12},  and single photon sources \cite{Ripka:18}. In addition, the use of thermal Rydberg atoms for electric field sensing is an exciting and growing field \cite{Sedlacek:12, Meyer:20, Berweger:23}, with microwave coupling of Rydberg states an important tool.

In order to connect quantum technologies based on microwave transitions with optical infrastructure, coherent microwave-to-optical conversion must be developed \cite{Lauk:20, Han:21}. One appealing possibility is the use of wave-mixing processes with microwave transitions.  Essential components of four-wave mixing through Rydberg states \cite{Brekke:08} and coherent microwave transitions between nearby Rydberg states \cite{Vogt:18} have been demonstrated.  These characteristics make  Rydberg atoms an appealing method for microwave-to-optical frequency conversion, with several methods currently being pursued \cite{Vogt:19, Gard:17, Covey:19, Tu:22, kumar:22, borówka:23, firdoshi:22}. Alternative Rydberg excitation schemes are being explored \cite{Thaicharoen:19, Fahey:11, Carr:12}, and hold appeal both for efficiency and in the interest of pursuing novel six-wave mixing schemes for microwave-to-optical conversion.

In this paper we explore Rydberg excitation in a rubidium vapor through an inverted excitation scheme, where EIT and fluorescent decay are used to investigate the process.  Resonant microwaves can couple nearby Rydberg states and cause Autler-Townes splitting, while detuned microwaves allow excitation directly to a secondary Rydberg state.  In our analysis, we observe the microwave AC stark shift near resonance, and demonstrate off resonant excitation for detunings from the intermediate Rydberg state of over 100 MHz.  This excitation detuning is sufficient to prevent unwanted excitation to the initial Rydberg level, and has the potential to be a fundamental component of six-wave mixing microwave-to-optical frequency conversion.    

\section{Experimental Setup}

\begin{figure}[ht!]
\centering\includegraphics[width=14cm]{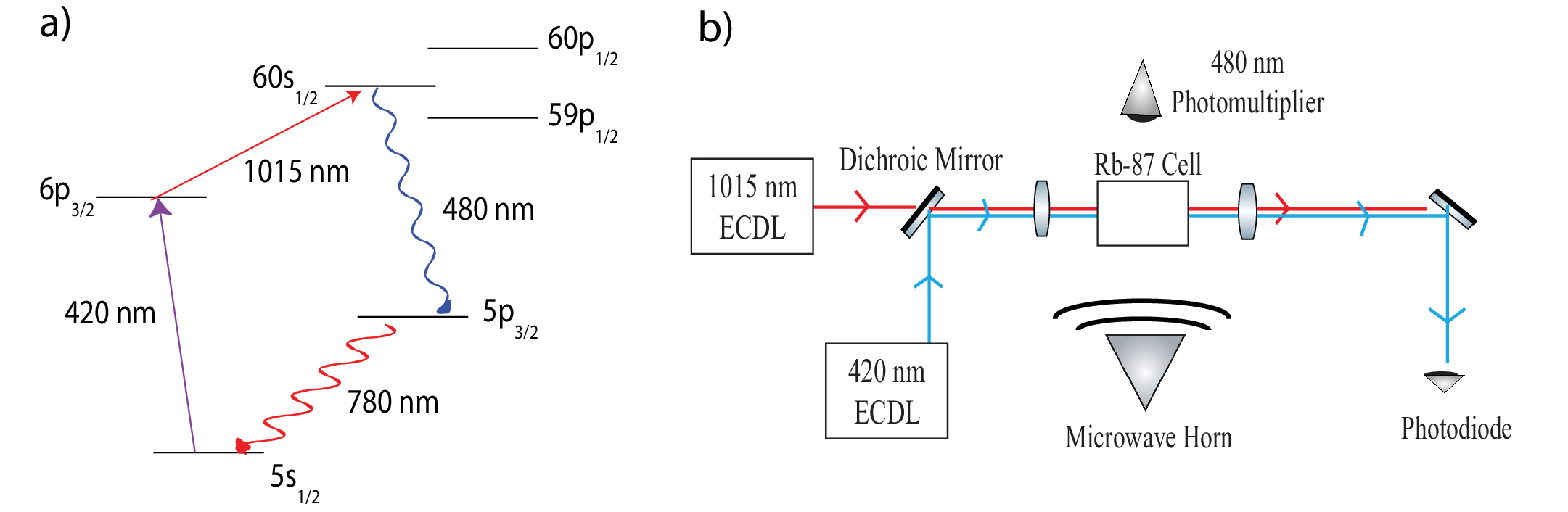}
\caption{a) The energy levels involved in Rydberg excitation and microwave coupling.  The initial 60s state can be microwave coupled to higher or lower np states at microwave frequencies.  b) A simplified version of the experimental setup.  }
\label{fig:expsetup}
\end{figure}

We excite rubidium 87 in a thermal vapor cell, with temperatures up to 125  $^{\circ}$C. A probe laser at 420 nm on the $5s_{1/2}$ to $6p_{3/2}$ transition is combined on a dichroic mirror with a coupling laser at 1015 nm on the $6p_{3/2}$ to $60s_{1/2}$ transition.  The energy levels involved can be seen in fig. \ref{fig:expsetup}a.  The 420 nm probe beam is focused to a waist of 15 $\mu$m and has Rabi frequency $\Omega_p=2\pi$ x 32 MHz, and the 1015 nm coupling beam is focused to a waist of 52 $\mu$m and has Rabi frequency $\Omega_c=2\pi$ x 11 MHz. A microwave horn perpendicular to the laser propagation direction can cause transitions between nearby Rydberg states.  

 In our setup the excitation scheme has the coupling laser with a higher wavelength than the probe laser. This inverted scheme results in much smaller EIT signals due to the one-photon and two-photon Doppler shifts having the same sign  \cite{Urvoy:13, Xu:15}. This limits the change in transmission with EIT to less less than 2\% in our case, but means the signal has similar strength with either co- or counter-propagating excitation lasers.  Here the lasers co-propagate in preparation for wave-mixing processes.  The transmission of the probe beam is observed on a photodiode, and the fluorescence at 480 nm on the $60s_{1/2}$ to $5p_{3/2}$ transition is detected on a photomultiplier.  The full experimental setup is seen in figure \ref{fig:expsetup}b.

\begin{figure}[ht!]
\centering\includegraphics[width=12cm]{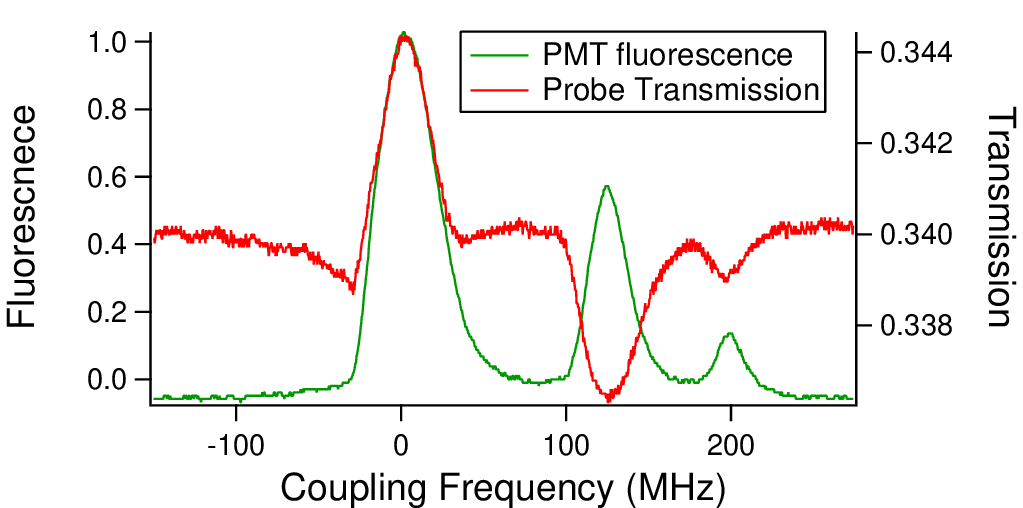}
\caption{ The fluorescence spectrum and probe transmission while the probe laser is on resonance and the coupling laser is scanned near the Rydberg resonance.  The 6p hyperfine structure is observed, with EIT on the $F'=3$ state.  }
\label{fig:spectrum}
\end{figure}

\section{Results and Analysis}

With the probe beam locked to the $5s_{1/2}$ to $6p_{3/2}$ transition, the coupling beam is scanned near the Rydberg resonance.  A spectrum showing the observed fluorescent decay and probe transmission is shown in fig. \ref{fig:spectrum}.  The $6p_{3/2}$ F$'=1,2,3$ hyperfine levels are observed.  As the beams are co-propagating, the hyperfine structure spacing is increased by a factor of $(1+\lambda_p/\lambda_c)$ due to velocity selection.  This spacing was used to determine the overall frequency scale of the scan, and confirmed with a Fabry-Perot cavity measurement of the coupling beam.  Despite the high cell temperatures, we are able to see the hyperfine structure with linewidths on order of 20 MHz, which may be limited by transit time broadening across the beam waist or collisional effects.   The spectrum is extremely sensitive to the waists and Rabi frequencies of the two beams \cite{Urvoy:13}, with the F$'=1,2$ states showing enhanced absorption.  

\begin{figure}[ht!]
\centering\includegraphics[width=14cm]{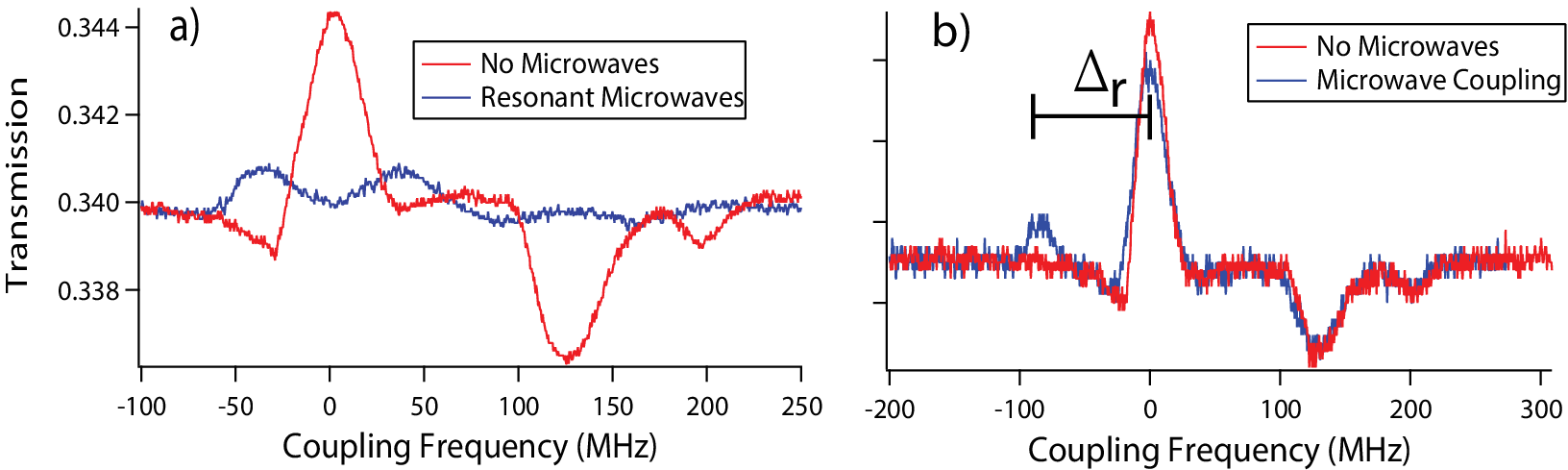}
\caption{ The probe transmission while the coupling laser is scanned near the Rydberg resonance with microwaves applied.  a) Microwaves on resonance with the $60s_{1/2} \rightarrow 60p_{1/2}$ transition results in Autler-Townes splitting. b) Microwaves detuned from the Rydberg-Rydberg transition by $\Delta_r$ producing a three photon excitation peak.   }
\label{fig:microwavescan}
\end{figure}

With microwaves on resonance with a nearby state, such as the $60s_{1/2}$ to $60p_{1/2}$ transition, Autler-Townes splitting can be observed in the spectrum, as seen in fig. \ref{fig:microwavescan}a.  From these splittings, we can observe that our microwave coupling strength produces Rabi frequencies of up to 70 MHz.  

As an alternative to on-resonant Rydberg excitation, we have also explored an off-resonant three photon excitation, which includes microwaves detuned from a Rydberg-Rydberg transition.  In this case, the excitation process $5s_{1/2} \rightarrow 6p_{3/2} \rightarrow 60s_{1/2} \rightarrow 60p_{1/2}$ is detuned a distance $\Delta_r$ from the the first $60s$ Rydberg state.  By detuning the coupling laser more than 50 MHz from resonance in the absence of microwaves, we can eliminate excitation to the Rydberg state, as observed by the lack of fluorescent decay.  The addition of off-resonant microwaves then causes excitation directly to the secondary Rydberg level.  This process can be observed by an additional off-resonant EIT signal created as the coupling laser is scanned, as shown in fig. \ref{fig:microwavescan}b.  

\begin{figure}[ht!]
\centering\includegraphics[width=14cm]{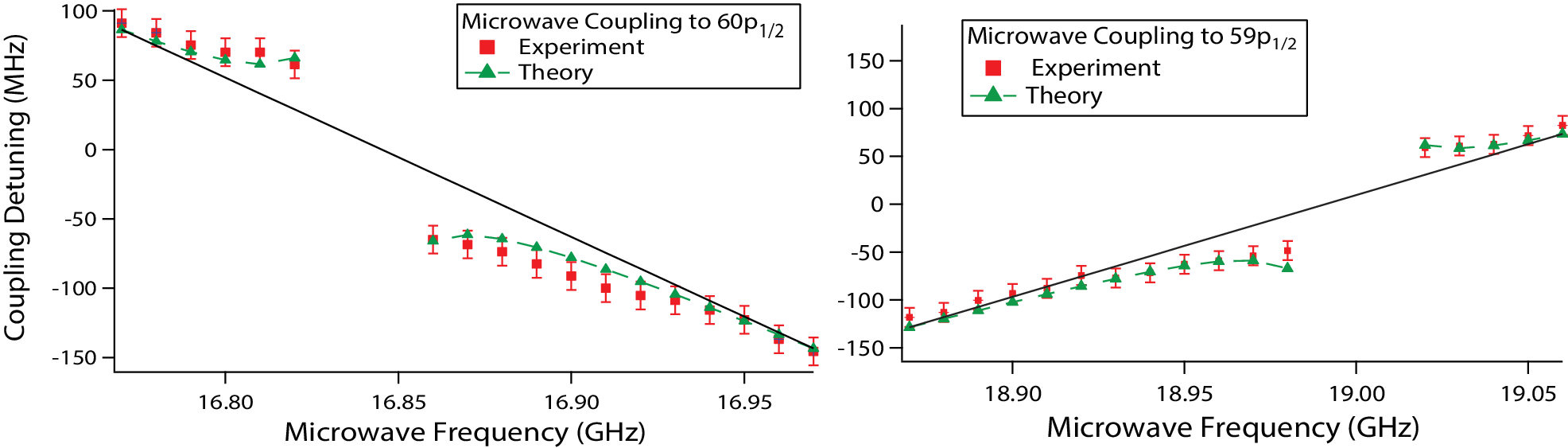}
\caption{ The detuning of the coupling laser required to obtain three photon excitation to the secondary Rydberg state for microwaves detuned from the Rydberg-Rydberg transition. Theory shows the AC stark shift from the microwaves causes a deviation from the proportional relationship near resonance.  The linear relationship is shown to guide the eye.  }
\label{fig:3photondualgraph}
\end{figure}

In order to investigate this off-resonant excitation process, we have varied the microwave frequency near the $60s_{1/2} \rightarrow 60p_{1/2}$ and $60s_{1/2} \rightarrow 59p_{1/2}$ transitions.  The change in coupling frequency needed to achieve off-resonant EIT is shown for each of these cases in fig. \ref{fig:3photondualgraph}.

Far from resonance, it is observed that the coupling laser detuning is linearly related to the microwave frequency.  The $60p_{1/2}$ state is above the original $60s_{1/2}$, so higher microwave frequencies require lower coupling frequencies to stay on resonant with the two-photon $6p_{3/2} \rightarrow 60p{1/2}$ transition, consistent with $\Delta_r/{\Delta_c}=-1$ far from resonance.  For the $ 59p_{1/2}$ state, where the secondary Rydberg state is below the original, this results in higher microwave frequencies leading to higher coupling frequencies, consistent with $\Delta_r/{\Delta_c}=1$ far from resonance.   Closer to resonance, the states are shifted by the AC stark shift due to the strong microwave Rabi frequency.  This is in agreement with previous experiments showing the microwave AC stark shift in Rydberg atoms could be used to shift resonance criteria \cite{Bohlouli:07}. For blue detuned microwaves where the states are shifted closer together, the 60p state is shifted down, while the 59p state is shifted up, with the opposite effect for red detuned microwaves.  This shift can be observed in the change in coupling frequency required for excitation near resonance shown in fig. \ref{fig:3photondualgraph}.  A simple theory of the AC stark shift approximated far from resonance as $\delta_{ac}=\Omega_{m}^2/4\Delta_r$ is fit to the data, revealing good agreement, and consistent with a microwave Rabi frequency of 60 MHz.

\section{Discussion}

In this paper we have demonstrated excitation to Rydberg states via a three-photon process including a detuned microwave transition.  This process has been observed through EIT on the probe beam, and shows that microwave coupling allows direct excitation to a secondary Rydberg state.  The shift of the excitation frequencies is consistent with the microwave-induced AC stark shift on Rydberg levels.  

The Rydberg excitation can be controlled by the presence of a microwave signal, and be significantly detuned from an initial Rydberg state. This excitation process is a useful step towards a variety of wave-mixing processes using Rydberg states, including the possibility of microwave-to-optical frequency conversion.



\begin{backmatter}
\bmsection{Funding}
NSF-RUI grant 2110357


\bmsection{Disclosures}
The authors declare no conflicts of interest.




\bmsection{Data availability} Data underlying the results presented in this paper are not publicly available at this time but may be obtained from the authors upon reasonable request.





\end{backmatter}



\bibliography{rydbergmicro}






\end{document}